\documentclass[11pt,twoside]{article}
\usepackage{./asp2014}

\aspSuppressVolSlug
\resetcounters

\begin{document}

\title{Clues to the Structure of AGN through massive variability surveys}
\author{Andy Lawrence, 
\affil{Institute for Astronomy, Univ. Edinburgh,\\ 
Scottish Universities Physics Alliance (SUPA)\\
Royal Observatory, Blackford Hill, Edinburgh, UK; \\
\email{al@roe.ac.uk}}}

\paperauthor{Sample~Author1}{Author1Email@email.edu}{ORCID_Or_Blank}{Author1 Institution}{Author1 Department}{City}{State/Province}{Postal Code}{Country}

\begin{abstract}
Variability studies hold information on otherwise unresolvable regions in Active Galactic Nuclei (AGN). Population studies of large samples likewise have been very productive for our understanding of AGN. These two themes are coming together in the idea of systematic variability studies of large samples - with SDSS, PanSTARRS, and soon, LSST. I summarise what we have learned about the optical and UV variability of AGN, and what it tells us about accretion discs and the BLR. The most exciting recent results have focused on rare large-scale outbursts and collapses - Tidal Disruption Events, changing-look AGN, and large amplitude microlensing. All of these promise to give us new insight into AGN physics.
\end{abstract}

{\em\noindent Extracted from "Astronomical Surveys and Big Data", 2016,\\
ASP Conf. Series, {\bf 505}, 109, eds. A.Mickaelian, A.Lawrence, T.Y.Magakian}

\section{AGN Survey Science}
Surveys have made a big difference to AGN science in three ways - through the systematic discovery of objects, through population analysis, and by the location of rare but important objects. A few examples will make the point.

(i) {\bf Systematic discovery.} It is particularly pleasing to give this talk in Byurakan, as this is where AGN science turned from the study of a handful of oddballs into a major component of modern astronomy. The Markarian objective prism surveys \citep{Markarian1967,Markarian1981} mark the beginning of systematic AGN studies. (ii) {\bf Population analysis.} The classic example of how we use complete samples statistically is the construction of the luminosity function - how AGN numbers are distributed with respect to luminosity and redshift. This is more than just dry statistics. We find that the total power output of the AGN population grew and then peaked a few billion years after the Big Bang, and has been declining since. Furthermore, less powerful AGN peaked in numbers at later times, a phenomenon known as ``cosmic downsizing''. The challenge is to explain these patterns in terms of the growth and fuelling of central black holes, in the context of galaxy evolution (see for example \citet{Hopkins2007}). (iii) {\bf Rare objects.} My favourite example of an important rare object is the quasar with the highest known redshift, ULASJ1120+0641, at z=7.085 \citep{Mortlock2011}. Finding this object required sifting through many millions of objects over thousands of square degrees. Again, this has more than just novelty value. It is a serious challenge to understand how to grow such a massive black hole only 700 million years after the Big Bang. The highest redshift quasars also constrain the ionisation history of the Intergalactic Medium.

We see these three themes - systematic discovery, population analysis, and rare objects - recur in the use of massive variability surveys. But first, what does variability itself tell us?

\section{AGN Variability Science}

All Type I AGN - those where we can see the strong blue continuum and broad emission lines - are variable.\footnote{For simplicity, I am only going to talk about the UVOIR spectral region, ignoring X-rays.} This is important because variability can provide indirect information on size scales that are otherwise unmeasurable. Suppose, for illustration, we take an AGN at a distance of 100 Mpc, and we assume that it contains a black hole of mass $10^8$ M$_\odot$. The Table below shows the angular scale of well known AGN structures, in units of the Schwarzschild radius $R_S=2GM/c^2$. The accretion disc, Broad Line Region (BLR) and the geometrically thick obscuring region sometimes known as the ``torus'' are all unresolvable by direct means, although as we will describe later, may be mappable by microlensing transits. 

If the accretion disc is in a stable steady state, we might expect it to evolve gradually on the inward drift timescale set by viscosity, which is of the order 10,000 years (see e.g. \citet{Netzer2013}). However, instabilities of various kinds could give us much faster changes. The {\em light crossing timescale} $t_{lt}=R/c$, is the shortest timescale that we could possibly see, if for example one region has variations locked to those of another region by radiation heating or reflection. This is of the order hours, days, and years for disc, BLR, and torus respectively. The {\em dynamical timescale}, $t_{dyn}=\sqrt{R^3/GM}$, is the shortest timescale on which we are likely to see physical changes in a region, and is of the order of days, years, and thousands of years for disc, BLR, and torus respectively. (Free-fall time is roughly the same and orbital timescale is $2\pi$ times longer.) More realistically, perturbations may transmit across a region on the sound crossing timescale $t_{snd}= R/ v_{snd}$. This is somewhat model dependent but is of the order of years for the accretion disc. Note what I mean here is the global time to cross the whole region. Local hot spots could grow on the timescale it takes sound to cross the vertical height of the disc, which could be 1--3 orders of magnitude faster. Somewhat related is the ``thermal'' timescale $t_{therm}$ which is roughly the time it takes for for energy to dissipate within the disc, i.e. it is a kind of response timescale to a spike of energy input. This is model dependent of course, but some standard formulae are given in \citet{Collier2001a} and \citet{Kelly2009}. It is of the order of days for the inner disc and years for the optical disc.  The analogous ``response'' timescale for the BLR and for the obscuring region is actually the light-crossing time - the local response time to a change in photo-ionisation or heating is very short, but what we see is smeared out by the range of light travel delays.

\begin{table}[!ht]
\begin{center}
\smallskip
{\small
\begin{tabular}{lllllll}  

\tableline
\noalign{\smallskip}
AGN & physical & angular & $t_{lt}$ & $t_{dyn}$ & $t_{snd}$ & $t_{therm}$ \\
\noalign{\smallskip}
Structure & size & size & &  &  & \\
\noalign{\smallskip}
\tableline

\noalign{\smallskip}
Inner disc & 5 $R_S$ & 0.1$\mu$as & 1.4hrs & 4.3hrs & 1.3 yrs & 18.7days \\
\noalign{\smallskip}
Optical disc & 50 $R_S$ & 1$\mu$as & 14hrs & 5.7days & 23 yrs & 1.6yrs \\
\noalign{\smallskip}
Broad Line Region & 1000 $R_S$ & 20$\mu$as & 11days & 1.4yrs & 800 yrs & -- \\
\noalign{\smallskip}
Obscuring Region & $10^5 R_S$ & 2mas & 3.1yrs & 1.4kyrs & 350 kyrs & --  \\

\noalign{\smallskip}
\tableline\
\end{tabular}
}
\end{center}
\end{table}

Are these timescales relevant to what we actually see? The UV continuum changes on timescales of weeks\footnote{Here I am assuming an Seyfert-like object appropriate to our $10^8 M_\odot$ example.}, with an RMS of around $\pm 30\%$, which means trough-to-peak changes of up to a factor of two are not unusual. The variations in the optical continuum, BLR, and IR seem to track these variations with roughly the light-travel time delays suggested in the Table, together with a similar amount of smearing (see recent examples in \citet{Edelson2015}, \cite{Grier2012}, and  \citet{Koshida2014}). This strongly suggests that almost all the changes we see on the relevant timescales represent reprocessed emission driven by changes in the very central regions. The conventional explanation for many years has been that the driving power is from the X-ray source (e.g \citet{McHardy2014}), but in many cases this does not work in either energy budget or correlation terms (see \citet{Lawrence2012} and references therein). A good alternative for the driving power is the (unseen) EUV peak of the very inner accretion disc.

The amplitude we see in the optical continuum on these $\sim$week timescales (around 3\%  RMS) is much smaller than that seen in the UV variations, which suggests that a very blue variable component mixes with an unchanging, or slower changing, redder component. \citet{Lawrence2012} argues that this variable reprocessor is a system of dense inner clouds surrounding the disc, rather than the disc itself.

The variations seen in the UV, which the optical and BLR emission track, seem to follow a red-noise or random-walk like pattern, increasing in amplitude to longer timescales, flattening at a characteristic timescale of the order tens of days. This timescale depends on the mass of the black hole (Collier and Peterson 2001). This characteristic timescale seems to match the thermal timescale of the inner disc, suggesting that variability is driven by some unknown stochastic process, filtered by the physical response of the disc \citep{Kelly2009, Kelly2011}.

Note that the changes we see in broad emission lines are also of the order weeks, tracking the changes in the UV photo-ionising source. This is much shorter than the dynamical timescale of the BLR, and means we are not seeing structural changes in this region. In the popular "local optimally emitting cloud (LOC)" models we will be lighting up different pre-existing clouds at different distances as the UV goes up and down \citep{Peterson2006,Goad2014}, which is why the amplitude of line variations (the ``responsivity'') varies with line species - Ly $\alpha$ has a large amplitude and Mg II hardly varies at all (e.g. \citet{Cackett2015}). However, it is possible that on longer timescales we {\em will} see BLR structural changes - a point we will return to in section 5.3.

\section{Massive variability projects}

The new frontier is massive variability studies - repeated surveying of large areas of sky. This concept was pioneered by Mike Hawkins and his multiple plates of UK Schmidt Field 287 \citep{Hawkins1993}, and then of course by the OGLE and MACHO projects, focused on looking for stellar microlensing events, but useful for much else. SDSS was not primarily a variability project, with the prominent exception of ``Stripe 82'' which is a 300 sq.degree strip which was observed many times \citep{Frieman2008}. Furthermore, because of the plate overlaps, a considerable fraction of the main SDSS area was in fact imaged twice, a fact which has been exploited by some workers (e.g. \citet{MacLeod2012}).

Over the last few years PanSTARRS-1 \citep{Kaiser2010} has imaged three quarters of the sky several times a year for three years (the $3\pi$ survey), along with a number of seven square degree deep fields imaged every few nights (the Medium Deep Surveys (MDS)). PanSTARRS-1 produced some very interesting results which we will describe below, but much data mining potential remains. 

All Type 1 AGN are variable. In principle, ongoing statistical variability could be used as the defining feature in sample selection \citep{Schmidt2010, MacLeod2011}, but of course many other astronomical objects are variable, so variability has sometimes been combined with colour selection \citep{Hawkins1993,Sarajedini2011, Peters2015,Palanque-Delabrouille2011}; see also AlSayyad et al in preparation. This seems to produce particularly clean samples of quasars, but so far has not produced really new types of object. It is a different story when we use repeated photometry to look for extreme variability and transients, whch has produced some fascinating surprises over the last few years. This is the story we will focus on in section \ref{sec:rare}. But meanwhile, what have massive variability studies told us about AGN variability in general?

\section{Massive variability projects: population studies}

The simplest use of large samples of objects is to make ensembles, improve statistics, and check the generality of results that were previously found for a handful of local Seyfert galaxies. For example, \citep{Welsh2011} use repeated GALEX observations to construct the UV structure function for $\sim$500 quasars. This confirms the red-noise like behaviour, a characteristic 30-40\% RMS, and a rollover at $\sim$100 days. (Note that these objects are more luminous than the local Seyfert galaxies, and so should have longer timescales.) At optical wavelengths, studies using Field 287, SDSS, POSS, and PanSTARRS confirm that the RMS variability is only $\sim 3-5\%$, but also that it keeps rising to longer timescales \citep{Hawkins2002,VandenBerk2004, DeVries2005, MacLeod2010, MacLeod2012, Morganson2014}. Whether or not there is definitely a turnover is somewhat controversial (compare \citep{DeVries2005}, \citep{MacLeod2010}, and \citep{Morganson2014}), but the characteristic timescale is certainly years rather than weeks, and the RMS does eventually reach the 30\% seen in UV variability. So it seems that the optical emission from quasars is just as variable as the UV emission, but takes longer to get there. How to square this with the UV-optical simultaneity seen in well sampled light curves of Seyfert galaxies is an important open question.

The classic Seyfert galaxy studies showed that variability amplitude depends on wavelength. This is strongly confirmed by the SDSS quasar studies \citep{VandenBerk2004, Schmidt2012}, which have the advantage that even broad-band filters can sample a range of rest wavelengths, because of the range of quasar redshifts. Fig. 13 of \citet{VandenBerk2004} is particularly striking, showing a smooth change of structure function amplitude from $\sim 0.15$mag at 6000\AA\ to $\sim 0.4$mag at 1000\AA\ .  

Several SDSS-based papers also discuss and demonstrate correlations of variability amplitude with redshift, luminosity, and estimated black hole mass \citep{VandenBerk2004,DeVries2005, MacLeod2010, MacLeod2012}. The correlation of variability amplitude with luminosity has a long history \citep{Hook1994, Giveon1999} and has also been seen in X-ray variability \citep{Lawrence1993,Almaini2000}. The luminosity effect, while very clear, is curiously weak. Fig. 11 in \citet{VandenBerk2004} shows that as the luminosity of a quasar ranges over a factor of 100, the variability amplitude changes only from $\sim 0.03$mag to $\sim 0.1$ mag. This is strongly inconsistent with simple ``shot noise'' style models, where variability is made up from multiple overlapping disconnected flares, and somewhat inconsistent with the simplest global variability models, where timescales should scale linearly with luminosity, which is indicating source size. How to explain the amplitude-luminosity effect is another important open question.

Another interesting feature comes from the study of \citet{MacLeod2012} who look at the distribution of two-epoch magnitude changes, and who find something with much broader high-amplitude tails than the naively expected Gaussian. Based on the \citet{MacLeod2010} fit to correlations with luminosity, redshift, and black hole mass, \citet{MacLeod2012} argue that the non-Gaussian distribution arises from superimposing Gaussians of a range of widths. 

Finally, we note an intriguing result from \citet{DeVries2005}, who show that the structure function amplitude is significantly larger for upward changes than downward changes, suggesting that variability may be constructed of sawtooth-like flares, but probably with a range of timescales and sizes. Despite the failure of simple fixed-flare shot-noise models to predict the luminosity-amplitude effect, more sophisticated versions may therefore still be of interest.

\section{Massive variability projects: rare objects}
\label{sec:rare}

\subsection{Tidal Disruption Events}

An exciting application of repeated imaging surveys is to spot the long-expected Tidal Disruption Events (TDEs), where a lone star passes close to a dormant black hole and is ripped apart, causing an accretion event that lasts a few months. If you can't wait a hundred thousand years, then you need to watch a hundred thousand galaxies in order to catch one going bang. Huge changes in luminosity that could correspond to such events were seen in X-ray and UV space-based surveys (e.g. \citep{Komossa1999, Komossa2004, Esquej2007, Gezari2008, Gezari2009}). The first systematic optical search was by \citet{VanVelzen2011} using the SDSS Stripe 82 data, who found two good examples. However, by far the most impressive and intriguing example so far has come from the PanSTARRS Medium Deep Survey \citep{Gezari2012}. The object concerned (PS1-10jh) flared by a factor of four hundred. Three years later, an HST observation shows a remaining UV source down by factor of 100 from the peak, centred on the nucleus of a small $z=0.16$ galaxy \citep{Gezari2015}. The dense time sampling and long monitoring meant both the rise and fall were well resolved for the first time, enabling a proper model fit. Even more important is that the rapid alert enabled a spectrum to be obtained at peak, as well as later in the decline. The rather bizarre result is that the flaring object was dominated by broad lines of HeII, with no sign of Hydrogen lines. It seems impossible to explain this by the flare-up of a pre-existing accretion disc, or by the accretion of an interstellar cloud. \citet{Gezari2012, Gezari2015} argue that the most plausible explanation is the partial disruption of the envelope of an evolved star which had already lost its hydrogen envelope.

\subsection{Slow-blue hypervariables}

\articlefigure[width=.5\textwidth, angle=-90]{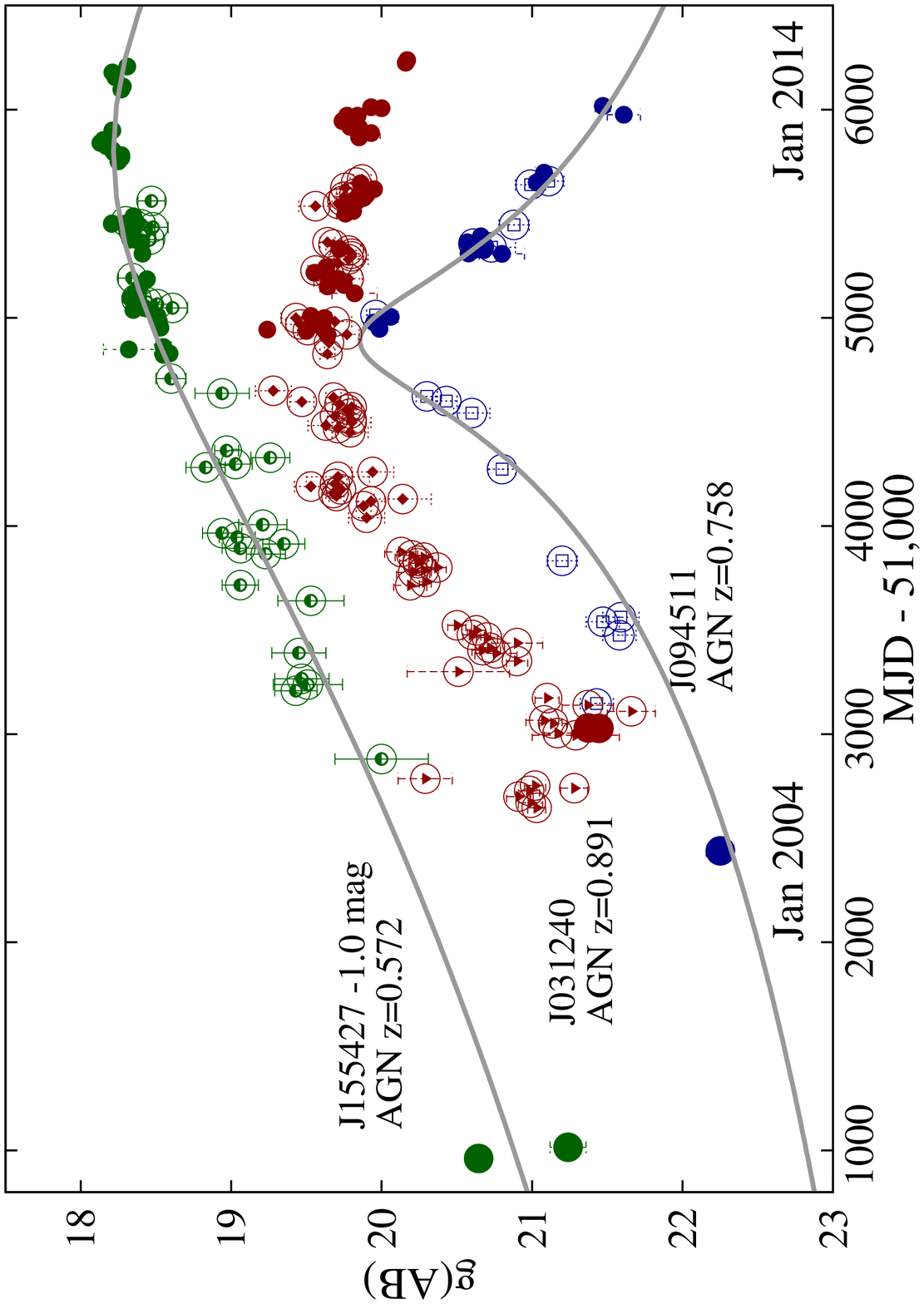}{fig:three-lc}{Light curves of slow-blue hypervariables from Lawrence et al 2016. Earliest solid symbols are SDSS photometry; later solid symbols are PanSTARRS or LT data; open symbols are heavily averaged CRTS data. The smooth curve through the J094511 data represents a simple microlensing model. The smooth curve through the J155427 data is not a model, but simply a smooth Bezier curve for illustration purposes. Figure taken from WHT proposal.}

Although the PanSTARRS search for TDEs initially focused on the frequently monitored MDS regions, we also looked in the less frequently monitored but much larger $3\pi$ survey, along with our colleagues hunting for supernovae.
 Because this work was started before a complete baseline sky image was ready, the search was performed as the PanSTARRS data came in by comparing pipeline photometry with the values for the corresponding objects a decade earlier in SDSS, and looking for changes of at least 1.5 magnitudes. This turned up several dozen examples of an unexpected new class of objects - slow, blue, nuclear hypervariables (Lawrence et al 2012, 2016). These were objects classified by SDSS as faint ($g\sim 21-23$) galaxies, but which by the PanSTARRS era had become point-like objects with $g\sim 19-20$. Follow-up monitoring with the Liverpool Telescope (LT) showed that some were red and fading on a timescale of weeks, but most were blue and evolving slowly, including some that were still rising. Follow-up spectroscopy with the William Herschel Telescope (WHT) showed that the red-fast objects were supernovae, and the slow-blue objects were AGN, typically at $z\sim 1$. Addition of archival data from the Catalina Real Time Transit Survey (CRTS, \citet{Drake2009}), showed us that these objects have undergone slow smooth symmetric outbursts by factors of many over ten years. (See Fig. \ref{fig:three-lc}.) These events are far too slow, and the time-integrated energy too large, to be TDEs; they could be accretion outbursts, or gravitational microlensing events caused by a star in an intervening galaxy (Lawrence et al 2016).

\articlefigure[width=.8\textwidth, angle=0]{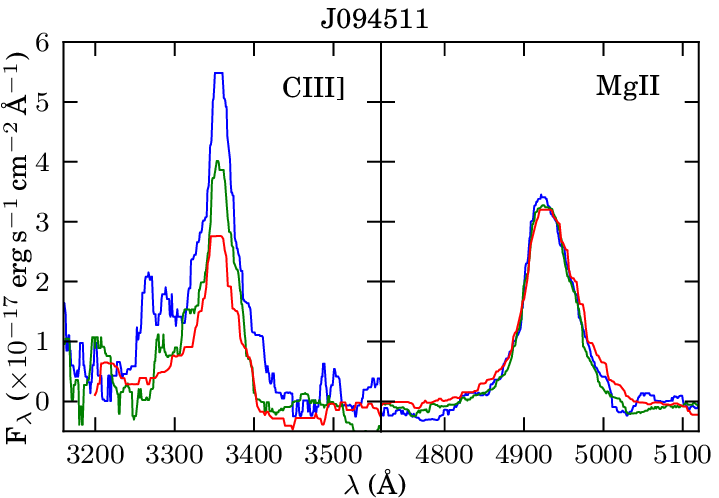}{fig:J094511-line}{Line profiles, from WHT spectra at three epochs during the declining portion of the J094511 light curve seen in Fig. 1. Mg II shows no change at all while CIII] shows both flux and profile changes. From Bruce et al in preparation.}

The timescales involved (years) are just what we expect for either possible structural changes in the BLR, or for the passage of the possible microlens across the BLR. In either case then, if we accumulate spectroscopic light curves, we could get significant diagnostic information about the BLR (especially if we can decide which of the two models is right!) We have been gradually accumulating such multiple epoch spectra. Initial results (Bruce et al in preparation) indicate that CIII] and CIV change by almost as much as the continuum, whereas MgII changes very little. If the outbursts are due to microlensing, our modelling suggests that the CIII] region is only a few light days across, whereas the MgII region is several times larger than this, and so suffers little amplification. Alternatively, if the outbursts are real accretion changes, the lack of response in Mg II suggests that we {\em still} see no structural changes in the BLR, even on timescales of years, which seems surprising.

\subsection{Changing look AGN}

We have known for many years that a minority of AGN can undergo large amplitude changes on a timescale of years (e.g. \citet{Tohline1976, Lawrence1977, Bischoff1999}). In some cases there has been a change in spectroscopic appearance from Type 2 or 1.8/1.9 to Type 1 \citep{Khachikian1971, Aretxaga1999, Shappee2014} or Type 1 to Type 2-ish \citep{Penston1984, LaMassa2015} or showing the appearance/disappearance of X-ray absorption \citep{Halpern1984, Matt2003, Risaliti2009, Marchese2012}. \citet{Matt2003} coined the term ``changing look AGN'' for these objects. At first the known examples were all local and relatively low luminosity. Recently however bulk searches in SDSS and PanSTARRS have found examples that are much more luminous, and so could reasonably be called changing-look quasars \citep{LaMassa2015, Ruan2015, MacLeod2016,Runnoe2015}. 

What causes these changes? Because many examples are very nearby, the outburst is unlikely to be due to line-of-sight microlensing. One possibility is a change in extinction. In general we believe that Type 2 AGN are Type 1 AGN with heavy extinction in the line of sight. This material will be in motion, so the line-of-sight extinction could easily change, especially in the context of ``clumpy torus'' models \citep{Nenkova2002}. Although the dynamical timescale, relevant to gross changes in the obscuring region, is of the order of thousands of years, the time for an occultation effect caused by a single  cloud to move across the face of the distant central source is far less, of the order of years. \citet{Goodrich1995} argues that changes are caused by motion of dust bearing clouds actually within the BLR. However, both \citet{LaMassa2015} and \citet{MacLeod2016} argue against extinction changes because we do not see the expected reddening signature. 

Alternatively, we may be seeing large changes in accretion rate, with the BLR responding to changes in the ionising continuum \citep{Shappee2014, LaMassa2015, MacLeod2016}. The timescale on which we see these dramatic changes (years) is similar to the ``knee timescale'' for optical changes for normal quasars discussed in section 4, but the amplitude of changes is far larger. The timescale concerned is possibly consistent with the accretion disc thermal timescale, but we don't have a theory for how such outbursts or collapses may happen. With respect to the BLR response, the \citet{MacLeod2016} paper contains an especially intriguing result; the quasar J0255+0230 shows a change in MgII flux as strong as the change in continuum. This is very different to the lack of MgII response we normally see in reverberation studies (see section 2). Either the nature of the changes is very different, or, because we are sampling a longer timescale (years as opposed to weeks) we are finally seeing the BLR respond physically, rather than just being ``lit up''.

Finally, we note that a small number of objects have been claimed over the years as ``true Type 2s'', i.e. Type 2 AGN that are not caused by extinction \citep{Tran2001, Panessa2002, Miniutti2013}, although the reality of some of these has been questioned by \citet{Stern2012}. It has been suggested that the disappearance of the BLR is a luminosity and/or evolutionary effect \citep{Elitzur2014}. However it is noticeable in the objects concerned that the Big Blue Bump is always absent at the same time as the broad lines, so it could well be that many of these objects are actually highly variable ``changing look' AGN which are only temporarily Type 2-like. Repeat spectroscopy of these objects could be important.

\section{Looking to the future}

The great opportunity of the next few years for massive variability studies is the Large Synoptic Survey Telescope (LSST; \citet{Ivezic2008}), an 8m-class telescope which will survey the whole southern sky repeatedly over ten years. It will measure the quasar variability structure function for huge numbers of individual objects rather than just ensembles, and together with spectroscopic followup, will enable a much more reliable analysis of how variability depends on AGN properties. LSST will also be ideal for catching rare transient and hypervariable objects, such as TDEs, changing look quasars, and microlensing events.

Ideally, what we {\em really} want next is a massive spectroscopy variability survey, especially on timescales of months to years, where it is becoming clear we are beginning to see structural changes in the BLR, and/or micro-lens transits across the BLR. Massive several-epoch spectroscopic studies have just started based on DES and SDSS quasars \citep{Yuan2015, Shen2015}, but these are focused on measuring reverberation. Another promising project is a subset of SDSS-IV known as the Time Domain Spectroscopic Survey (TDSS; \citet{Morganson2015}), which will classify 220,000 variable objects and take repeated spectra of a subset. Of what course what we would {\em really} like is a dedicated degree-scale spectroscopic monitoring machine. This would be  a thing of great joy to all of us old LAGS (Lovers of Active Galaxies).

\acknowledgements I would like to thank Areg Mickaelian for organising a wonderful and diverse meeting, and my Edinburgh colleagues, Chelsea MacLeod, Alastair Bruce, and Nic Ross, for lively, provocative, and helpful discussions.

\bibliography{lawrence_a_refs}  

\end{document}